\documentclass{kluwer}
\usepackage[dvips]{graphicx}

\newcommand{\SN}{{_{\rm SN}}}

\newcommand{\SNII}{{_{\rm SN\,II}}}

\newcommand\la{\;
  \raise0.3ex\hbox{$<$\kern-0.75em\raise-1.1ex\hbox{$\sim$
  }}\;\hskip-2pt }
\newcommand\ga{\;
  \raise0.3ex\hbox{$>$\kern-0.75em\raise-1.1ex\hbox{$\sim$
  }}\;\hskip-2pt }

\newcommand{\cm}{\,{\rm cm}}

\newcommand{\cmsq}{\,{\rm cm^{-2}}}
\newcommand{\cmcube}{\,{\rm cm^{-3}}}
\newcommand{\dyn}{\,{\rm dyn}}

\newcommand{\g}{\,{\rm g}}

\newcommand{\kms}{\,{\rm km\,s^{-1}}}

\newcommand{\K}{\,{\rm K}}
\newcommand{\kpc}{\,{\rm kpc}}

\newcommand{\mkG}{\,\mu{\rm G}}

\newcommand{\yr}{\,{\rm yr}}
\newcommand{\Myr}{\,{\rm Myr}}

\begin{document}
\begin{article}
\begin{opening}

\title{The effects of spiral arms on the multi-phase ISM}
\author{Anvar \surname{Shukurov} \& Graeme R.\ \surname{Sarson}}
\institute{School of Mathematics \& Statistics, University of
Newcastle, Newcastle NE1 7RU, U.K.}
\author{{\AA}ke \surname{Nordlund}}
\institute{NBIfAFG \& TAC,
Juliane Maries Vej 30, DK-2100 Copenhagen \O, Denmark}
\author{Boris \surname{Gudiksen}}
\institute{Inst.\ Solar Physics, Royal Swedish Acad.\ of Sciences,
SE-106 91 Stockholm, Sweden}
\author{Axel \surname{Brandenburg}}
\institute{NORDITA, Blegdamsvej 17, DK-2100 Copenhagen \O, Denmark}

\runningauthor{A.\ Shukurov et al.}
\runningtitle{The effects of spiral arms on the multi-phase ISM}
\date{\today}

\begin{abstract}
Statistical parameters of the ISM driven by thermal energy injections
from supernova explosions have been obtained from 3D, nonlinear,
magnetohydrodynamic, shearing-box simulations for spiral arm and
interarm regions. The density scale height obtained for the interarm regions
is 50\% larger than within the spiral arms because of the
higher gas temperature. The filling factor
of the hot gas is also significantly larger between the arms
and depends sensitively on magnetic field strength.
\end{abstract}
\keywords{ISM, turbulence, galactic spiral structure}

\end{opening}

%-----------------------------------------------------------
Detailed observations of the multi-phase interstellar medium (ISM)
have recently been supplemented with extensive numerical simulations
that capture much of the physics involved, including
driving by supernovae (SNe) and radiative cooling.
Several of the
models are in 3D (e.g., de Avillez, 2000;de  Avillez and Berry, 2001;
de Avillez and Mac Low, 2001), but only that of Korpi et
al.\ (1999a,b) includes rotation and so is suitable for
realistic modelling of magnetic fields. Here we discuss a
further development of this model, where we concentrate on the effects
of the total gas column density and magnetic field strength on the
properties of the ISM.

%-----------------------------------------------------------
Our models for regions inside and outside the spiral arms differ in
the gas column density; the midplane density $\rho_0$ is
given in Table~I.
The three models with low, intermediate
and high densities are referred to as Interarm, Average and Arm.
The column density does not vary in time in our simulations. Thus,
our model is applicable not far from
the corotation radius of the spiral pattern, and so is complementary to
the model of G\'omez \& Cox (2002) which focuses on dynamic effects
due to the passage of a density wave through the ISM.

%----------------------------------------------------------
\begin{table}[tb]                        \label{table1}
\begin{tabular}{lcccc}
\hline
                                        &Unit  &Interarm &Average & Arm \\
Initial midplane density,
        $\rho_{0}$ &$10^{-24}\g\cmcube$ & 0.7   & 1.4   & 2.9   \\
\hline
Density Gaussian scale height &$\!\kpc$         & 0.23  & 0.20  & 0.16  \\
Mean temperature     &$10^4\K$                  & 35    & 7.8   & 3.4  \\
Mean thermal pressure & $10^{-14}\dyn\cmsq$     & 42    & 68    & 120  \\
Rms vertical velocity &$\!\kms$                 & 23    & 20    & 20   \\
Mean turbulent pressure &  $10^{-14}\dyn\cmsq$  & 39    & 63    & 110  \\ 
Hot gas filling factor     &                    & 0.12  & 0.07  & 0.04  \\
SN II rate &${\rm kpc^{-2}}\,{\rm Myr}^{-1}$    & 11    & 38    & 111   \\
SN II Gaussian scale height &kpc                & 0.30  & 0.16  & 0.14  \\
\hline
\end{tabular}
\caption{Three models with varying initial mid-plane density,
$\rho_{0}$.
The mean temperature, pressures, root mean square velocity and filling factor
are all calculated within $|z| < 0.2 \kpc$.
The filling factors are for $T>10^{5}\K$; 
the initial magnetic field at the midplane is $B_{0} = 6 \mkG$ in all models.}
\end{table}
%----------------------------------------------------------

\begin{figure}[t]
\leftline{\includegraphics[width=2.25in]{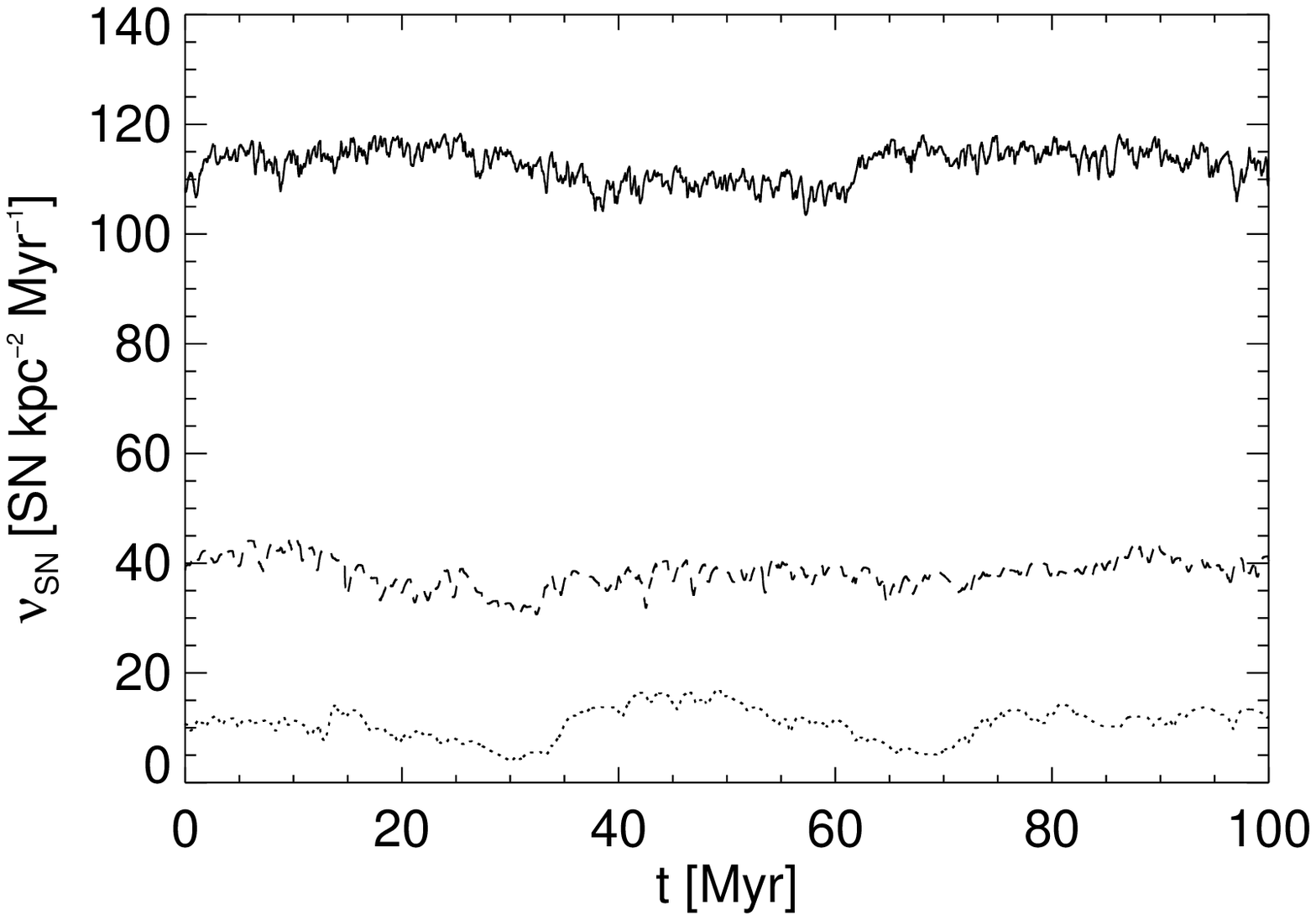}}
\vspace*{-41mm}
\rightline{\includegraphics[width=2.25in]{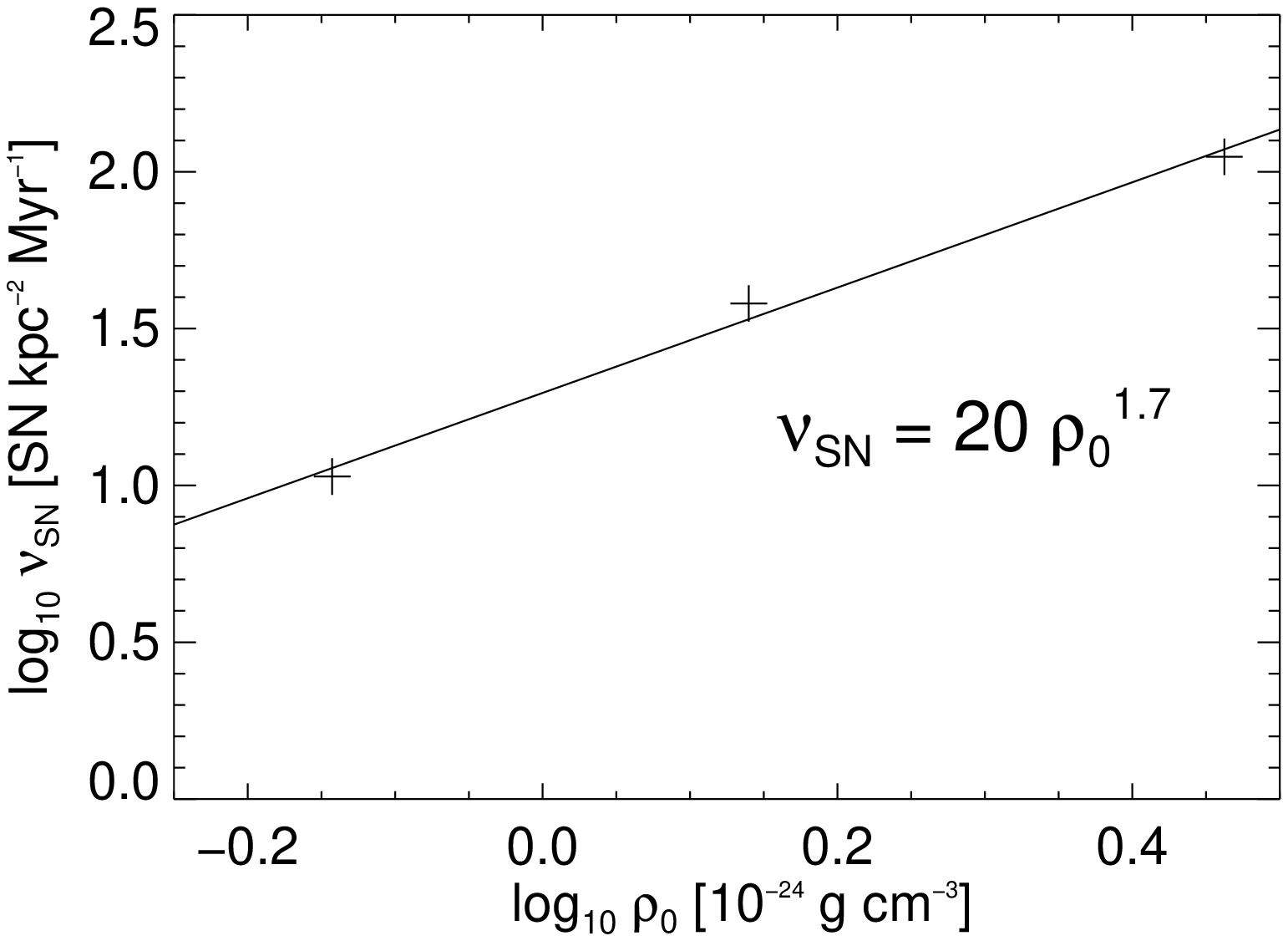}}
\caption{The SN II rate versus time (left), for the Arm (solid),
Average (dashed) and Interarm (dotted) models.
The scaling of the mean SN II rate with mid-plane density, $\rho_{0}$ (right).}
\label{snrate}
\end{figure}

We solve numerically the 3D, non-ideal
MHD equations with rotation, density stratification in external
gravity, heat sources and radiative cooling (Korpi et
al., 1999a). The simulations start with an imposed azimuthal
magnetic field of variable strength $B_0$, with vertical profile
$B_0\cosh^{-2}(z/0.3\kpc)$.
The system is driven by localised thermal energy injections,
modelling SN explosions. Type II SNe are initiated at those
locations where the gas density $\rho$ exceeds
$\rho_{\rm c}=10^{-24}\g\cm^{-3}$ and temperature $T$ is lower than
$T_{\rm c}=4000\K$; the probability of an SN explosion at a
given position is proportional to the local gas density within those
regions.  The SN~II rate, $\nu\SNII$, depends on the local ISM parameters via
(Gudiksen, 1999)
\[
\nu\SNII=\frac{M_{\rm c} X\SN X_*}{M\SN\tau_{\rm c}}\;,
\]
where $M_{\rm c}$ is the mass of gas where $\rho>\rho_{\rm c}$ and
$T<T_{\rm c}$, $X\SN=0.1$ is the fraction of the stellar
mass in SN progenitors obtained from a suitable initial mass function,
$X_*=0.02$ is the gas mass fraction converted into stars,
$M\SN=10\,M_\odot$ is the SN progenitor mass, and $\tau_{\rm
c}=20\Myr$ is the gas recycling time. The latter has been
adjusted to have $\nu\SNII\simeq40\kpc^{-2}\Myr^{-1}$ ($1/33\yr$ for
the whole Galaxy) for the {\em average\/} gas density; we stress that
both the rate and spatial distribution of Type II SNe are not prescribed
but rather follow from the simulations. The resulting dependence of
the SN~II rate on the mean gas density, shown in Fig.~\ref{snrate},
is consistent with the observed scaling of the star formation rate with
gas surface density, ${\rm SFR}\propto\Sigma^\kappa$ with
$\kappa=0.9$--1.7 (Kennicutt, 1998). The SN~II rates
and scale heights are given in Table~I.
Our model also includes Type I SNe implemented as in Korpi et al.\ (1999a).

%----------------------------------------------------------
%----------------------------------------------------------

%----------------------------------------------------------
\begin{figure}[b]
\leftline{\includegraphics[width=2.25in]{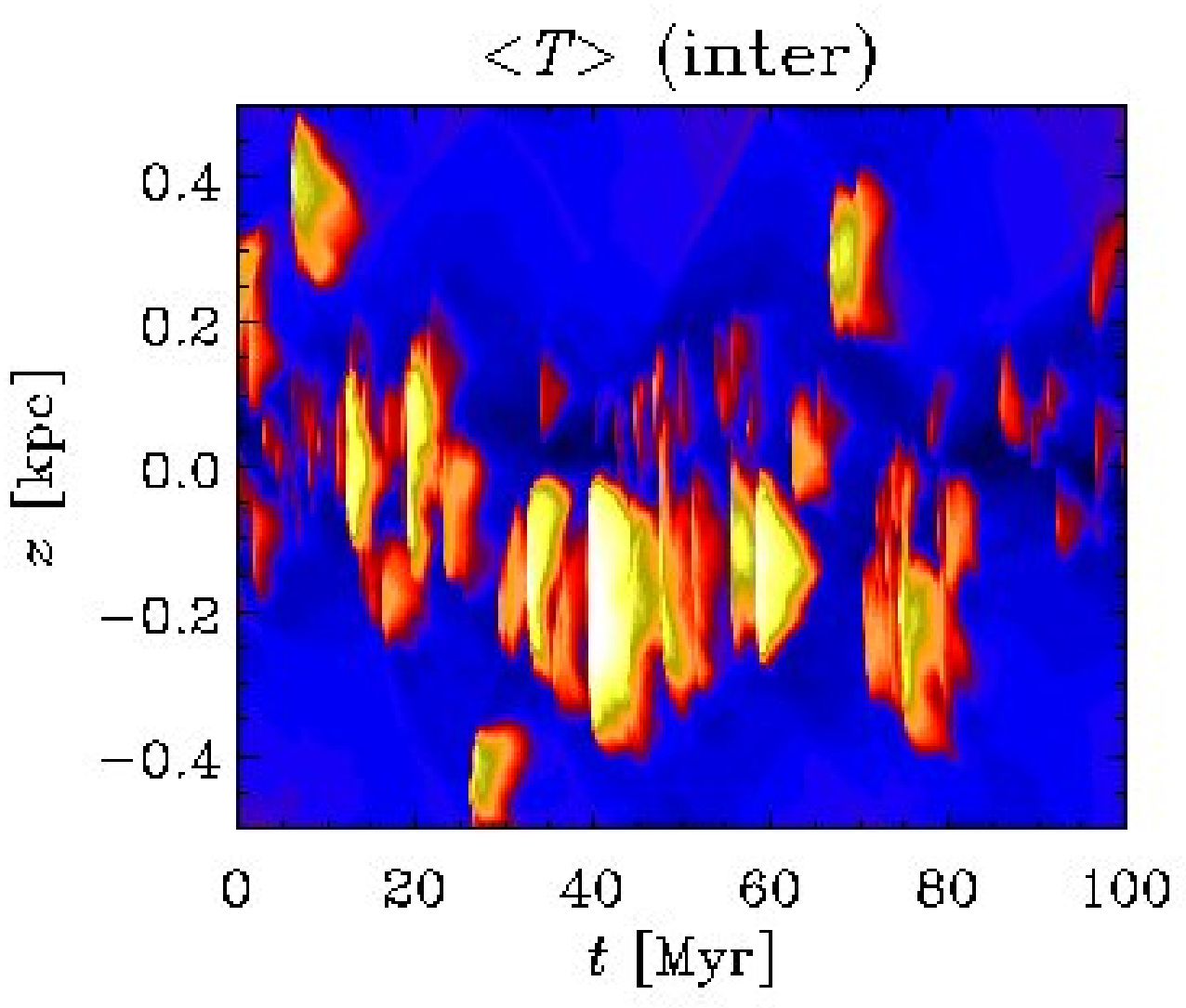}}
\vspace*{-46mm}
\rightline{\includegraphics[width=2.25in]{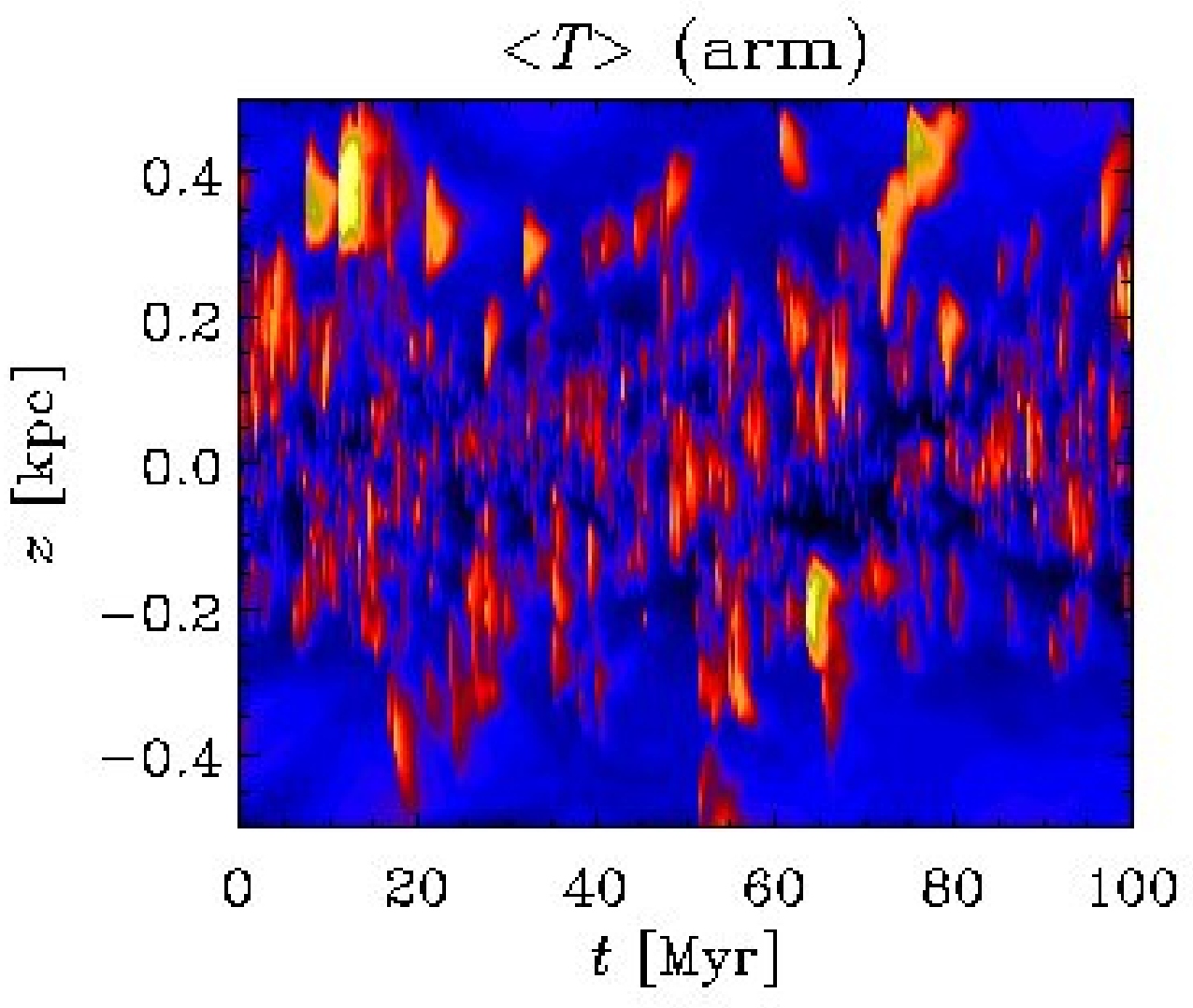}}
\caption{Horizontally averaged temperature as a function of height and time,
for the Interarm (left panel) and Arm (right panel) solutions. Lighter shades of grey
indicate higher temperatures.}
\label{TTm}
\end{figure}
%----------------------------------------------------------

Our results, as reported here, are preliminary because we use a
relatively small computational domain with the horizontal and
vertical ($z$) dimensions of  $0.25\times0.25\times1\kpc$ (with the
midplane in the centre), modest spatial resolution of about 4\,pc, and
closed boundary conditions in $z$. Furthermore, the cooling function
was truncated at $500\K$ to avoid high densities. All these
restrictions are of a technical nature and can readily be relaxed.

An unexpected result of our simulations is that the density scale
height is significantly larger in the Interarm model, although
both thermal and turbulent pressures are a factor of about 3 larger
in the Arm model.
The reason is that the gas temperatures are
higher in the Interarm case (Fig.~\ref{TTm}),
even though the SN rate is lower.
An apparent reason is that the cooling rate has a stronger
net dependence on gas density than the SN energy injection rate. 

Another surprising feature of the results presented in
Table~I is that the filling factor of the hot gas is lower than expected
by a factor of 2--3. This can be attributed to the geometry of the
magnetic field in our models;  it is uniform initially,
and therefore effective in confining expanding bubbles of hot gas. The
initial midplane field strength, $6\mkG$, is close to that of the {\em
total\/} field in the Solar vicinity, but the field is
implausibly well ordered. A more realistic simulation
would initialize the model with a ratio of turbulent to ordered
magnetic energy of about 3. We postpone such calculations for future
work, but here we explore the dependence of the results on the
strength of the initially uniform magnetic field; the
results are shown in Table~II. The filling factor of the
hot gas is sensitive to the field strength and increases to
0.2 as the field becomes weaker.
The density scale height marginally increases
with magnetic field strength,
but this effect is much less pronounced than the suppression of the hot phase;
the field strongly suppresses turbulence in the hot gas.
Thus, the magnetic field can affect the disc-halo
connection and the global structure of the ISM in
crucial, diverse and unexpected ways. This
aspect of the ISM dynamics has not yet been fully explored.

%-------------------------------------------------------
\begin{table}[t]                \label{table2}
\begin{tabular}{lccc} \hline
Initial magnetic field strength,
$B_{0}$                 &$\mkG$              &  0       & 6    \\
\hline
Density scale height    &kpc                  &  0.20  & 0.23 \\
Mean thermal pressure & $10^{-14}\dyn\cmsq$   &  50    & 42   \\
Rms vertical velocity &$\!\kms$               &  43    & 23   \\
Mean turbulent pressure & $10^{-14}\dyn\cmsq$ &  54    & 39   \\
Hot gas filling factor  &                     &  0.19  & 0.12 \\
\hline
\end{tabular}
\caption{The effect of magnetic field on the ISM illustrated with two
runs with varying initial magnetic field, $B_{0}$.
All variables are as defined in Table~I.
All runs have $\rho_{0} = 0.7 \times 10^{-24}\g\cmcube$ (the Interarm model).}
\end{table}
%---------------------------------------------------------
%---------------------------------------------------------------
\acknowledgements
We are grateful to M.-M.~Mac Low for helpful comments on the
dependence of the filling factor of the hot gas on magnetic field.
Use of the Copenhagen branch of the Danish Center for Scientific Computing
and the PPARC supported UKAFF computer facility at Leicester
is acknowledged.
This work was supported by the PPARC Grant PPA/G/S/2000/00528.

\theendnotes

\end{article}
\end{document}